\documentclass{iopart}
\input epsf

\def\labela{\def\theequation{\arabic{equation}a}}
\def\labelb{\addtocounter{equation}{-1}\def\theequation{\arabic{equation}b}}
\def\labelc{\addtocounter{equation}{-1}\def\theequation{\arabic{equation}c}}
\def\labeld{\addtocounter{equation}{-1}\def\theequation{\arabic{equation}d}}
\def\labelz{\def\theequation{\arabic{equation}}}

\begin{document}


\title{Exciton entanglement in two coupled semiconductor microcrystallites}

\author{Yu-xi Liu$^1$, \c{S}ahin K. \"Ozdemir$^{1,2,5}$,
Adam Miranowicz$^{1,2,3}$, Masato Koashi$^{1,2,5}$, and Nobuyuki
Imoto$^{1,2,4,5}$}

\address{$^1$ The Graduate University for Advanced Studies
(SOKENDAI), Shonan Village, Hayama, Kanagawa 240-0193, Japan}

\address{$^2$ SORST Research Team for Interacting Carrier
Electronics, Hayama, Kanagawa 240-0193, Japan}

\address{$^3$ Nonlinear Optics Division, Physics Institute, Adam
Mickiewicz University, 61-614 Pozna\'n, Poland}

\address{$^4$ NTT Basic Research Laboratories, 3-1
Morinosato-Wakamiya, Atsugi, Kanagawa 243-0198, Japan}

\address{$^5$ CREST Research Team for Photonic Quantum Information,
Hayama, Kanagawa 240-0193, Japan}

\begin{abstract}\\
Entanglement of the excitonic states in the system of two coupled
semiconductor microcrystallites,  whose  sizes are much larger
than the Bohr radius of exciton in bulk semiconductor but smaller
than the relevant optical wavelength, is quantified in terms of
the entropy of entanglement.  It is observed that the nonlinear
interaction between excitons increases the maximum values of the
entropy of the entanglement more than that of the linear coupling
model. Therefore, a system of two  coupled microcrystallites can
be used as a good source of entanglement with fixed exciton
number. The relationship between the entropy of the entanglement
and the population imbalance of two microcrystallites is
numerically shown and the uppermost envelope function for them is
estimated by applying the Jaynes principle.

\vspace{3mm} \noindent {\sc published in J. Phys. A  {\bf 37},
4423-4436 (March 2004)}

\noindent {PACS numbers 03.67.-a, 71.10.Li, 71.35.-y, 73.20Dx}

\end{abstract}

\pagenumbering{arabic}
\section{Introduction}

Recently, quantum computation and quantum information excite large
enthusiasm among theoretical and experimental physicists. Many
theoretical and experimental researches have been devoted to the
preparation and measurement of the entangled states that have been
considered to be a key ingredient in the realization of the
quantum computation and information.

We can quantitatively  understand  the role of quantum
entanglement as a resource for communication by studying the
quantum teleportation, which was first realized for discrete
variables~\cite{Zeilinger} and then for continuous
variables~\cite{B}.  The possibility for the generation of the
maximally entangled states with a fixed photon number from
squeezed vacuum states or from mixed Gaussian continuous entangled
states by the quantum non-demolition measurement~\cite{mmm} has
been theoretically studied. Quantum teleportation using an
entangled source of fixed total photon number has also been
theoretically investigated in~\cite{c}. However, it has been shown
that a generation of the maximally entangled states with a fixed
finite photon number is a great challenge to modern technology. So
the exploration of new maximally entangled bipartite source with
fixed particle number is very interesting and important  from both
experimental and theoretical view points.

Several  schemes based on coupled quantum dots have been proposed
for fabricating quantum gates~\cite{4,44}. The entanglement of the
exciton states in a single quantum dot  or in a  quantum dot
molecule has  been demonstrated experimentally \cite{5,6}. The
authors of references \cite{7,99,yu-xi2,wang} theoretically
investigated the entanglement of excitonic states in the system of
the optically driven coupled quantum dots and propose models to
prepare  Bell, Greenberger-Horne-Zeilinger (GHZ) and $W$ states.
The above investigations mainly focus on the case in which the
sizes of quantum dots  are smaller than the Bohr radius of exciton
in bulk semiconductor, so that the quantum dots have discrete
energy levels and one energy state can only be occupied by one
exciton because of the Pauli principle. In such quantum dot
systems, it is easy to define a qubit by different spins of
excitons, or by single-exciton and no-exciton states. So when we
want to encode quantum information in systems of quantum dots by a
finite-dimensional quantum state space (called qudit system), we
need to consider higher energy levels. However, if we can enlarge
the size $R$ of the quantum dot so that it is larger than the Bohr
radius $a_{B}$ of the exciton in the bulk semiconductor but
smaller than the relevant light wavelength $\lambda$, in such a
system of quantum microcrystallites (which can also be called
large quantum dots system~\cite{Han,Ban,Eng}), the excitons can
occupy the same level when the average number of excitons per Bohr
radius volume is less than one. Then we can encode quantum
information in the finite Hilbert space built by the different
exciton number states. Quantum information properties of
semiconductor microcrystallites coupled by a cavity field were
investigated by us in Refs. \cite{yu-xi} where, in particular, we
have proposed a realization of symmetric sharing of entanglement
for the exciton states in semiconductor microcrystallites and also
studied the environment effect on the qubit.

Motivated by these considerations, in this paper we will study a
system of two  semiconductor microcrystallites coupled by Coulomb
interaction. We will study the entanglement of the qudit excitonic
states in these two coupled semiconductor microcrystallites with
fixed exciton number.

The paper is organized as follows: In Sec. II, we
 first give a theoretical description of two coupled
semiconductor microcrystallites. We obtain the analytical solution
of the system by virtue of the Schwinger representation of the
angular momentum. In Sec. III, the entanglement of the two
subsystems is discussed using the von Neumann entropy under
certain initial conditions. We discuss and  numerically  analyze
the oscillation of the exciton population imbalance between two
quantum dots. The relationship between the entanglement and
oscillation of the imbalance population of excitons is
demonstrated in detail in Sec. IV. Finally, we give our
conclusions and some comments.

\section{The model and its analytical solution}

We consider a system which consists of two  semiconductor
microcrystallites  coupled by Coulomb interaction. We  assume that
two semiconductor microcrystallites are completely symmetric.
Their sizes are larger than the Bohr radius $a_{B}$ of an exciton
in a bulk semiconductor but smaller than the relevant light
wavelength $\lambda$. Using two-band approximation to model these
two coupled semiconductor microcrystallites, we have the
Hamiltonian
\begin{equation}
H=\hbar\sum_{j,k=1}^{2}\chi_{jk}a^{\dagger}_{j}a_{k}+\sum_{j,k,l,m=1
}^{2} \chi_{jklm}a^{\dagger}_{j}a^{\dagger}_{k}a_{l}a_{m},
\label{eq:Ham}
\end{equation}
where $a^{\dagger}_{j}\,(a_{j})$ are creation (annihilation)
operators of  excitons, which are electron-hole pairs bound by the
Coulomb interaction. We assume that the density of the excitons is
so low and the external confinement potential to microcrystallite
is so weak that the exciton operators $a^{\dagger}_{j}\,(a_{j})$
can be described approximately as bosonic operators,  that is they
satisfy the commutation relations of the ideal bosons, $[a_{j},
a^{\dagger}_{k}]=\delta_{jk}$, which is somewhat different from
that in Ref.~\cite{d}. The label $j$=1 (2) denotes the
microcrystallite $A\,(B)$. In the Hamiltonian (\ref{eq:Ham}), the
deviation of the exciton operators from the ideal bosonic model is
corrected by introducing an effective nonlinear interaction between
the hypothetical ideal bosons. In general, the parameters
$\chi_{jk}$ are different  from each other, however in this study
we consider two  completely equivalent  microcrystallites, which
have nearly the same Bohr radius of the excitons and transitional
dipole moment. For the sake of simplicity, we assume the parameters
$\chi_{jk}=\omega
>0$ for $j=k$, which means that the two semiconductor
microcrystallites  have the same transition frequency from the
valence band to the conduction band, and assume  positive real
numbers $\chi_{jk}=g$ for $j \neq k$, which correspond to a linear
coupling constant of the two semiconductor microcrystallites. The
parameters $\chi_{jklm}$ of the nonlinear terms are also taken as
positive constant and set to be $\chi_{jklm}=\chi$.   Under these
assumptions, the Hamiltonian (\ref{eq:Ham}) can be simplified as
follows
\begin{eqnarray}
H&=&\hbar\Omega {\bf N}+\hbar\chi {\bf N}^{2}
+\hbar{\bf G}(a^{\dagger}_{1}a_{2}+a^{\dagger}_{2}a_{1})\nonumber \\
&&+\hbar\chi(a^{\dagger}_{1}a_{2}+a^{\dagger}_{2}a_{1})^{2},
\label{eq:ham}
\end{eqnarray}
where  $\Omega=\omega-2\chi$, and ${\bf G}=g-2\chi+2\chi {\bf N}$.
It is obvious that ${\bf N}=a^{\dagger}_{1}a_{1}
+a^{\dagger}_{2}a_{2}$ is a constant of motion, which means $[{\bf
N},H]=0$ and the total exciton number of the coupled semiconductor
microcrystallites  is conserved. In such a situation, we find that
it is more convenient to get the solution of the Schr\"odinger
equation governed by Hamiltonian (\ref{eq:ham}) by virtue of
Schwinger representation of the angular momentum
\cite{Sch65,Sak94}. That is, we can introduce the angular momentum
operators \labela
\begin{eqnarray}
J_{x}&=&\frac{1}{2}(a^{\dagger}_{1}a_{2}+a^{\dagger}_{2}a_{1}),\label{eq:J1}
\end{eqnarray}
\labelb
\begin{eqnarray}
 J_{y}&=&\frac{1}{2i}(a^{\dagger}_{1}a_{2}-a^{\dagger}_{2}a_{1}),
\label{eq:J2}
\end{eqnarray}
\labelc
\begin{eqnarray}
J_{z}&=&\frac{1}{2}(a^{\dagger}_{1}a_{1}-a^{\dagger}_{2}a_{2})\label{eq:J}
\end{eqnarray}
\labelz%
from the bosonic annihilation and creation operators of
the two exciton modes. The operators of
(\ref{eq:J1})--(\ref{eq:J}) satisfy the commutation relations for
the Lie algebra of SU(2):
\begin{equation}
[J_{j}, J_{k}]=i\varepsilon_{jkl} J_{l}, \hspace{0.5cm}
j,k,l=x,y,z,
\end{equation}
where the Levi-Civit\`a tensor $\varepsilon_{jkl}$ is equal to
$+1$ and $-1$ for the even and the odd permutation of its indices,
respectively, and zero otherwise. From
(\ref{eq:J1})--(\ref{eq:J}), the total angular momentum operator
can be expressed as
\begin{equation}
J^{2}=\frac{\bf N}{2}(\frac{\bf N}{2}+1).
\label{eq:Ca}
\end{equation}
In fact, ${\bf N}$ itself commutes with all the operators
(\ref{eq:J1})--(\ref{eq:J}) and (\ref{eq:Ca}). For a fixed total
excitonic number $\cal{N}$, the common eigenstates of $J^{2}$ and
$J_{z}$ are the two-mode Fock states \cite{Sch65,Sak94}
\begin{equation}
|j,m\rangle_s=|m_{1},m_{2}\rangle=
\frac{(a_{1}^{\dagger})^{j+m}(a_{2}^{\dagger})^{j-m}}{\sqrt{(j+m)!(j-m)!}}
|0\rangle \label{eq:schwinger}
\end{equation}
with eigenvalues $j={\cal N}/2$ and $m=-{\cal N}/2, \cdots, {\cal
N}/2$, where $|m_{1},m_{2}\rangle$  is the Fock  state with
$m_{1}=j+m$ excitons and $m_{2}=j-m$ excitons in  microcrystallite
$A$ and microcrystallite  $B$ respectively. Although $j\pm m$
$(m_{1}, m_{2})$ must be integers, $j$ and $m$ can both  be
integers or half-odd integers. For consistency, $j$ is replaced by
${\cal N}/2$ in the following expressions. For clarity,  the
subscript $s$ is used to indicate the Schwinger angular momentum
basis in (\ref{eq:schwinger}) and the following equations. By using
equation (\ref{eq:schwinger}), we take $j=1$ and $m=0$ as a
detailed example to show the relation between the Schwinger angular
momentum basis and Fock state basis. That is,
$|1,0\rangle_{s}=a_{1}^{\dagger}a_{2}^{\dagger}|0\rangle=|1,1\rangle$
which means that when the quantum number of the total angular
momentum $j$ is $1$ and its $z$ component $m$ is $0$, there is one
exciton in each microcrystallite respectively.

In terms of an ${\rm SO(3)}$ rotation $e^{i{\pi}/{2}J_{y}}$ of
$2\hbar{\bf G}J_{x}+4\hbar \chi J_{x}^{2}$, equation (\ref{eq:ham})
can be simplified into :
\begin{eqnarray}
H&=&\hbar \Omega {\bf N}+\hbar\chi{\bf N}^{2}+2\hbar {\bf
G}J_{x}+4\hbar \chi
J_{x}^{2}\nonumber \\
&=&\hbar\Omega {\bf N}+\hbar\chi{\bf N}^{2}+2\hbar {\bf G}
e^{-i({\pi}/{2})J_{y}}J_{z}e^{i({\pi}/{2})J_{y}}\nonumber \\
&&+4\hbar \chi e^{-i({\pi}/{2})J_{y}}J_{z}^{2}
e^{i({\pi}/{2})J_{y}}. \label{eq:h2}
\end{eqnarray}
The eigenfunctions $\Psi_{{\cal N}/2,m}$ and the eigenvalues
$E_{{\cal N}/2,m}$ of Hamiltonian (\ref{eq:h2}) can be obtained
easily as
\labela%
\begin{eqnarray}
\label{eq:8a}|\Psi_{\frac{\cal N}{2},m}\rangle_{s}&=&e^{(-i\pi/2)
J_{y}}|\frac{\cal N}{2},m\rangle_s
 =\sum_{m^{\prime}=-\frac{\cal
N}{2}}^{\frac{\cal N}{2}}{\cal D}^{\frac{\cal N}{2}}_{m^{\prime}m}
({\textstyle\frac{\pi}{2}})|\frac{\cal N}{2} ,m^{\prime}\rangle_s,
\end{eqnarray} \labelb%
\begin{eqnarray} E_{{\cal N}/2,m}&=&\hbar\Omega {\cal
N}+\hbar\chi{\cal N}^2 +2\hbar {\cal G} m+ 4\hbar \chi m^{2},
\end{eqnarray}
\labelz%
where ${\cal N}$ denotes the total excitonic number of two
microcrystallites; ${\cal G}=g-2\chi+2\chi {\cal N}$, and Wigner's
formula for ${\cal D}^{\frac{\cal
N}{2}}_{m^{\prime}m}(\frac{\pi}{2})$ reads as
\begin{eqnarray}
{\cal D}^{\frac{\cal
N}{2}}_{m^{\prime}m}(\frac{\pi}{2})&=&\sum_{k}(-1)^{k-m-m^{\prime}}
(\textstyle\frac{1}{2})^{\frac{\cal N}{2}} \nonumber \\
&&\times \frac{\sqrt{(\frac{\cal N}{2}+m)!(\frac{\cal
N}{2}-m)!(\frac{\cal N}{2}+m^{\prime})!(\frac{\cal
N}{2}-m^{\prime})!}} {(\frac{\cal N}{2}+m-k)!k!(\frac{\cal
N}{2}-k-m^{\prime})!(k-m+m^{\prime})!}, \label{eq:wg}
\end{eqnarray}
where we take the sum over $k$ so that none of the arguments of
factorials in the denominator is negative. The wave function
$|\Psi (t)\rangle$ of the system for the initial condition
$|\Psi(t=0)\rangle$ is given by
\begin{eqnarray}
|\Psi(t)\rangle&=& \exp(-i t H/\hbar) |\Psi (0)\rangle=\sum_{{\cal
N}=0}^{\infty}\sum_{m=-{\cal N}/2}^{m={\cal N}/2}\nonumber\\
&&\times\exp\left(\frac{E_{\frac{\cal N}{2}, m}}{i\hbar}t \right)
\langle\Psi_{\frac{\cal N}{2}, m}|\Psi(0)\rangle |\Psi_{\frac{\cal
N}{2}, m}\rangle_{s}. \label{eq:T}
\end{eqnarray}
The coefficients $\langle\Psi_{\frac{N}{2}, m}|\Psi(0)\rangle$ are
rotating matrix elements that can be determined by Wigner's
formula. Equation (\ref{eq:T}) is basic and will be used in our
further discussions. In the following  two sections, we will
discuss the entanglement of two exciton modes and demonstrate the
oscillations of population imbalance for excitons between two
quantum microcrystallites.

\section{Entanglement of the excitonic states}

Quantum entanglement plays the key role in the quantum information
and quantum computation.  In general, for any pure composite state
$|\psi\rangle_{AB}$ of  a bipartite system whose state space is
$H_{A}\otimes H_{B}$, the entanglement can be measured by von
Neumann's entropy of any reduced density operator $\rho_{A}={\rm
Tr}_{B}(|\psi\rangle\!_{AB}\,_{AB}\!\langle\psi|)$ or
$\rho_{B}={\rm Tr}_{A}(|\psi\rangle\!_{AB}
\,_{AB}\!\langle\psi|)$, where the reduced density operator of
system $A$ is obtained by tracing out system $B$ and that of
system   $B$  by tracing out system $A$. In the context of quantum
computation and information, in particular in studies of quantum
entanglement, the reduced von Neumann entropy of a bipartite pure
state $|\psi\rangle_{AB}$ is usually referred to as the {\em
entropy of entanglement} or simply the {\em entanglement} (see,
e.g., \cite{Ben96a,Ben96b,Ved98,Pho91}):
\begin{eqnarray}
E(\rho)&=&-{\rm Tr}(\rho_{A}\ln\rho_{A})=-{\rm
 Tr}(\rho_{B}\ln\rho_{B})
 =-\sum_{n}\lambda_{n}\ln\lambda_{n} \label{eq:ss}
\end{eqnarray}
where $\lambda_{n}$ are the eigenvalues of either $\rho_{A}$ or
$\rho_{B}$. They form the (square of the) coefficients of the
Schmidt decomposition of the  bipartite pure state. That is, the
bipartite pure state  $|\psi\rangle_{AB}$ can be expressed by a
set of  biorthogonal  vectors using the Schmidt decomposition as
\begin{equation}
|\psi\rangle\equiv
|\psi\rangle_{AB}=\sum_{n}\sqrt{\lambda_{n}}|\alpha_{n}\rangle_{A}
|\beta_{n}\rangle_{B} \label{eq:ab}
\end{equation}
where we have chosen the phases of our basis states so that no
phases appear in the coefficients $\lambda_{n}$ in the sum of
(\ref{eq:ab}), $\{|\alpha_{n}\rangle, n=0 \cdots \}$ and
$\{|\beta_{n}\rangle,n=0\cdots \}$ are orthonormal states of the
two subsystems $A$ and $B$  respectively.

The total number of excitons in the whole system is fixed by the
initial condition, such as $L$. Based on such a condition, the
maximally entangled state of the system is
\begin{equation}
|M\rangle=\frac{1}{\sqrt{L+1}}\sum_{l=0}^{L}|L-l, l\rangle,
\label{eq:M}
\end{equation}
where $|L-l,l\rangle$ represents the state with $L-l$ excitons in
 microcrystallite $A$ and $l$ excitons in microcrystallite $B$.
According to definition (\ref{eq:ss}), the  entropy of the
entanglement for the maximally entangled state (\ref{eq:M}) is
$\ln(1+L)$.

Experimentally, the easier way is to excite some number of
excitons in one microcrystallite, or both of microcrystallites.
Without loss of generality, we first assume that microcrystallite
$A$ is initially excited with $L$ excitons, and no excitons
initially exist in microcrystallite $B$. So the modes of the two
microcrystallites are disentangled at the initial time $t=0$. That
is, the state of the whole system is a tensor product of the
states of two subsystems $A$ and $B$, i.e. $|\Psi(t=0)\rangle
=|L\rangle_{A}|0\rangle_{B}$  with the Schwinger realization of
this initial state as  $|\Psi(t=0)\rangle
=|\frac{L}{2},\frac{L}{2}\rangle_s$. The system and each mode of
the system are in pure states and have zero entropies.

It is well known that any pure state remains pure with the unitary
evolution, but it is not true for each subsystem. With the
evolution,  the initially pure state of each subsystem can be
transformed into mixed states. If the von Neumann entropy
$E(\rho)$, given by Eq. (\ref{eq:ss}), is increased,  then the two
subsystems become more entangled. Thus,  the degree of
entanglement between the two subsystems in the system at any time
can be described by (\ref{eq:ss}).

\begin{figure}
\hspace{15mm} \epsfxsize=12cm\epsfbox{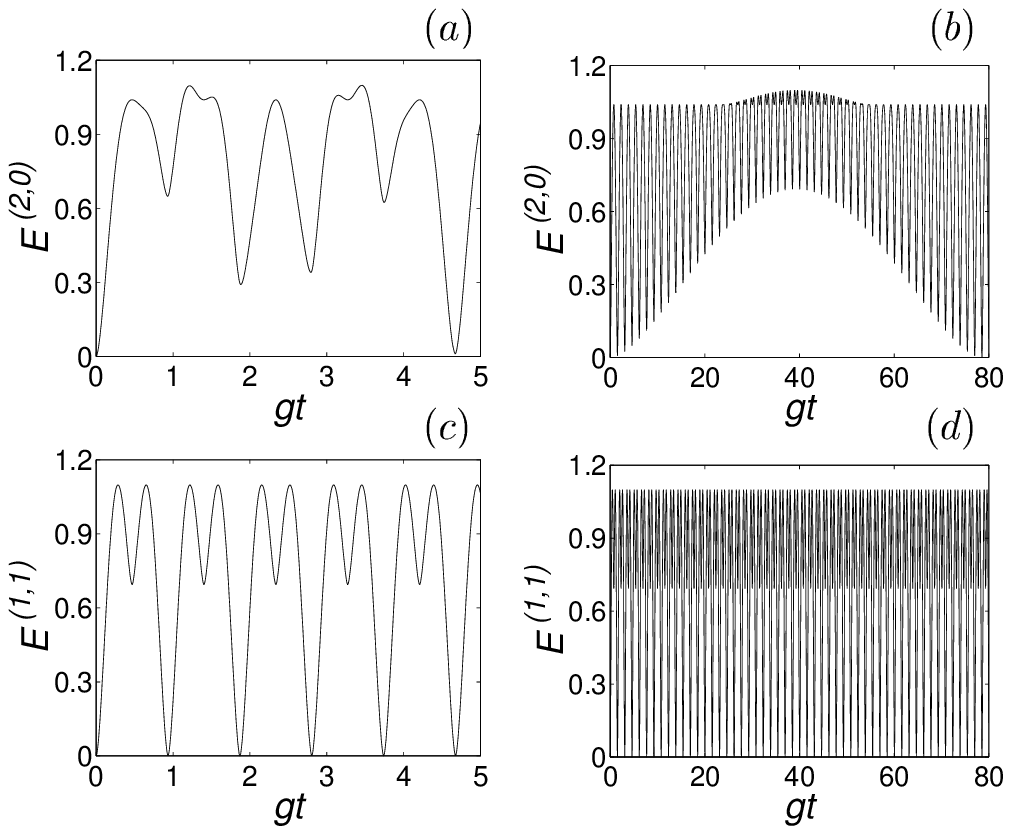}
\caption[]{$E^{(2,0)}$ is  plotted as a function of $gt$ for (a)
$\chi/g=0.34$, (b) $\chi/g=0.01$  when two excitons are initially
excited in one of the microcrystallites,  and $E^{(1,1)}$ is
plotted as a function of $gt$ for (c) $\chi/g=0.34$, (d)
$\chi/g=0.01$  when there is one exciton initially in each
microcrystallite.}\label{fig1}
\end{figure}

\begin{figure}
\hspace{15mm} \epsfxsize=12cm\epsfbox{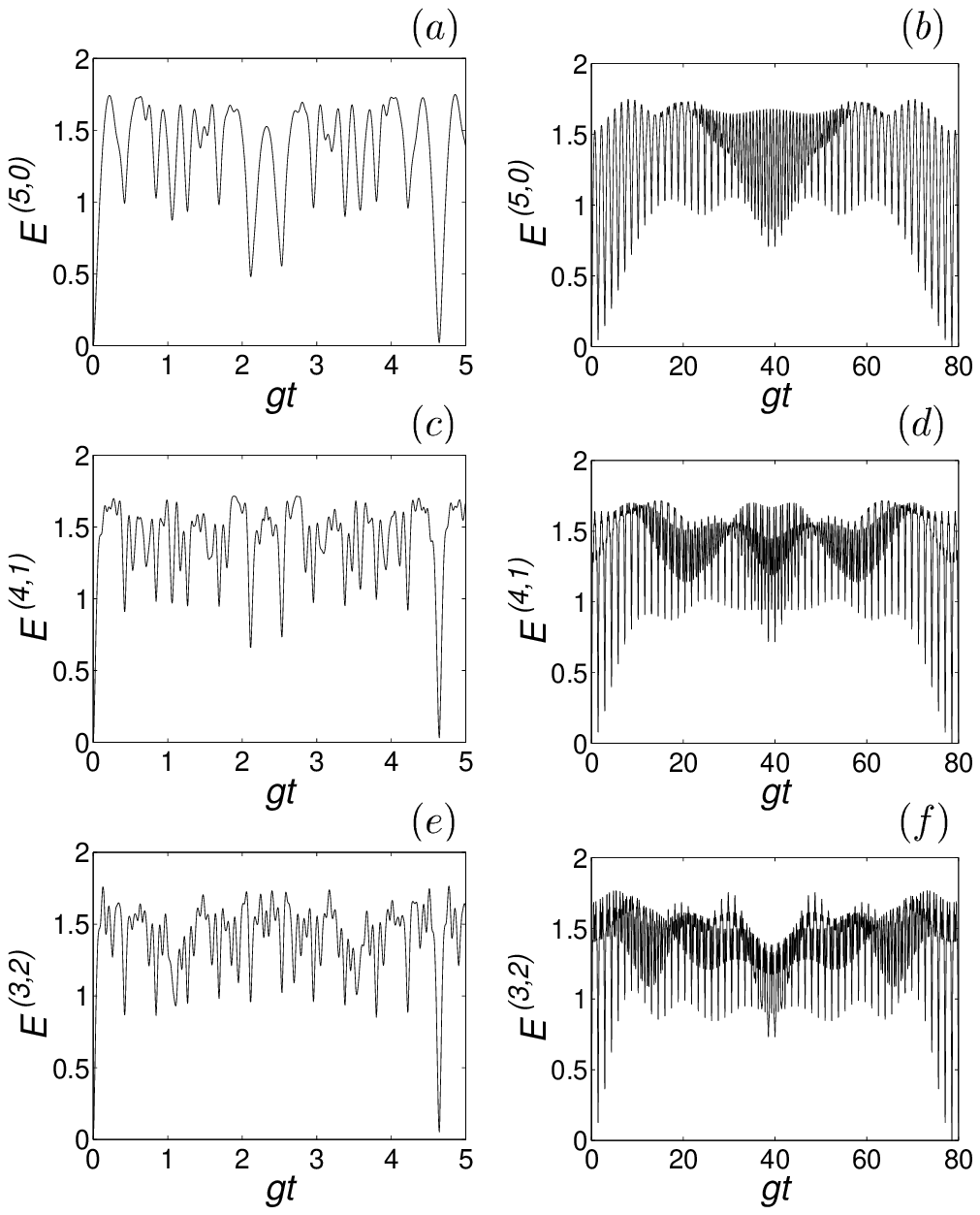}
\caption[]{$E^{(5)}\equiv E^{(5,0)}$, $E^{(4,1)}$, and $E^{(3,2)}$
are plotted as a function of $gt$ for $\chi/g=0.34$ (left panel)
and $\chi/g=0.01$ (right panel).}\label{fig2}
\end{figure}

For the initial state $|\frac{L}{2},\frac{L}{2}\rangle_s$ of the
system, the total wave function of the system can be obtained from
(\ref{eq:T}) as
\begin{equation}
|\Psi(t)\rangle=\sum_{l^{\prime}=-L/2}^{l^{\prime}=L/2}
\exp\Big(\frac{E_{\frac{L}{2},l^{\prime}}}{i\hbar}t\Big)
\langle\Psi_{\frac{L}{2},l^{\prime}}|\frac{L}{2},\frac{L}{2}\rangle_s
|\Psi_{\frac{L}{2},l^{\prime}}\rangle_{s}, \label{eq:13}
\end{equation}
where the normalized coefficients
$\langle\Psi_{\frac{L}{2},l^{\prime}}|\frac{L}{2},\frac{L}{2}\rangle_s$
are determined by  (\ref{eq:8a})-(\ref{eq:T}). We will use equation
(\ref{eq:13}) to discuss  the degree of the entanglement of the two
subsystems for the following several concrete examples. First, we
consider that there is initially one exciton in the
microcrystallite $A$, i.e., $L=1$. In this case, we can obtain the
wave function of the whole system from (\ref{eq:13}) and
(\ref{eq:schwinger}) with the initial state
$|\frac{1}{2},\frac{1}{2}\rangle_s$ as
\begin{eqnarray}
|\Psi^{(1)}(t)\rangle&=& \cos(gt)|\frac{1}{2},\frac{1}{2}\rangle_s
-
i\sin(gt)|\frac{1}{2},-\frac{1}{2}\rangle_s\nonumber \\
&=&\cos(gt)|1\rangle_{A}|0\rangle_{B}- i
\sin(gt)|0\rangle_{A}|1\rangle_{B}, \label{eq:nn}
\end{eqnarray}
where subscripts $A$ and $B$ denote dots $1$ and $2$,
respectively, and we have omitted the global phase factor of time
dependence, $e^{-i(\Omega+2\chi)t}$. The entropy of entanglement
can be calculated as
\begin{equation}
E^{(1)}(t)=-{\cos}^{2}(gt)\ln[\cos^{2}(gt)] -{\sin}^{2}(gt) \ln
[\sin^{2}(gt)]. \label{eq:nnn}
\end{equation}
It is seen that the entropy of the entanglement periodically
evolves with zero values at times $gt=k \frac{\pi}{2}$ where $k$
is an integer. The entropy of the entanglement reaches its maximum
values $\ln2$ at times $gt=(2k+1)\frac{\pi}{4}$ with integer $k$,
 then the maximally entangled states of the system is
\begin{eqnarray}
|\Psi^{(1)}(t)\rangle&=& \frac{|\frac{1}{2},\frac{1}{2}\rangle_s
+e^{-i 2gt}|\frac{1}{2},-\frac{1}{2}\rangle_s}{\sqrt{2}}
=\frac{|1\rangle_{A}|0\rangle_{B} +e^{-i
2gt}|0\rangle_{A}|1\rangle_{B}}{\sqrt{2}}.
\end{eqnarray}
If there are initially two excitons in microcrystallite $A$, then
the wave function of the whole system with the initial condition
$|\Psi(t=0)\rangle=|1,1\rangle_s$ becomes
\labela%
\begin{eqnarray}
|\Psi^{(2)}(t)\rangle &=& \alpha_{1}|1,-1\rangle_s
+\alpha_{2}|1,0\rangle_s+\alpha_{3}|1,1\rangle_s\nonumber
\nonumber \\
&=&\alpha_{1}|0\rangle_{A}|2\rangle_{B}+\alpha_{2}
|1\rangle_{A}|1\rangle_{B}+\alpha_{3}|2\rangle_{A}|0\rangle_{B}
\end{eqnarray}
and
\labelb%
\begin{eqnarray}
\alpha_{1}&=&\frac{1}{4}[e^{i2gt}+e^{-i2(g+4\chi)t}-2],
\label{eq:x1}
\end{eqnarray}
\labelc%
\begin{eqnarray}
\alpha_{2}&=&\frac{\sqrt{2}}{4}[e^{-i2(g+4\chi)t}-e^{i2gt}],
\end{eqnarray}
\labeld%
\begin{eqnarray}
\alpha_{3}&=&
\frac{1}{4}[e^{i2gt}+e^{-i2(g+4\chi)t}+2]\label{eq:x3}
\end{eqnarray}
\labelz%
where the global phase factor  of  time dependence $e^{-i 2(\Omega
+2\chi)t}$  has also been neglected. We  can obtain the entropy of
the entanglement as
\begin{equation}
E^{(2)}(t)=-\sum_{n=1}^{3}|\alpha_{n}|^{2}\ln|\alpha_{n}|^{2},
\end{equation}
where $\alpha_{n}$ determined by (\ref{eq:x1})--(\ref{eq:x3}).
There is no solution for the time $t$ in which the system exactly
evolves into a maximally entangled state such that the entropy of
entanglement is $\ln 3$. But numerically we find that the
difference between the maximum value of the entropy of the
entanglement and the entropy of the maximally entangled state is
of the order $10^{-5}$.  At this point, we say we can nearly
obtain the maximally entangled state when the total number of the
excitons in the system is equal to two. Figures \ref{fig1} and
\ref{fig2}  demonstrate the evolution of the entropy of the
entanglement as a function of $gt$ with two different sets of
parameters for given different initial states. They show that when
the second-order coupling between excitons is weak, the period for
reaching the maximum values of the entropy of the entanglement
becomes longer. It is worth pointing out that if the second-order
coupling constant is equal to zero and only linear coupling term
remains, then the exciton density is lower and the entanglement is
diminished (see solid curve with dots in comparison to other
curves in figure 3). So, in order to obtain high entanglement of
excitons, we should increase the exciton density in the coupled
microcrystallites. The above discussion also shows that the
entanglement between two microcrystallites depends on both the
ratio $\chi/g$ and initial conditions. By numerical calculations,
we can find that the survival time of higher entanglement also
depends on  $\chi/g$ and initial conditions. The survival time of
higher entanglement is longer for the larger ratio $\chi/g$ when
there are two excitons initially excited in one of the
microcrystallites, but it almost the same for different ratios
$\chi/g$ when there is one exciton initially excited in each
microcrystallite, respectively.

For an arbitrary number of excitons $L$, the wave function of the
whole system is described by (\ref{eq:13}). Then the entropy of the
entanglement can be calculated as
\labela%
\begin{equation}
E^{(L)}(t)\equiv E^{(L,0)}(t)=
-\sum_{m^{\prime}=-\frac{L}{2}}^{\frac{L}{2}}
|\beta_{m^{\prime}}|^{2}\ln|\beta_{m^{\prime}}|^{2} \label{eq:g1}
\end{equation}
with
\labelb%
\begin{equation}
\beta_{m^{\prime}}=\sum_{m=-\frac{L}{2}}^{\frac{L}{2}} {\cal
D}^{\frac{L}{2}}_{\frac{L}{2},m}(\frac{\pi}{2})
e^{-\frac{i}{\hbar}E_{\frac{L}{2},m}t} {\cal
D}^{\frac{L}{2}}_{m^{\prime},m}(\frac{\pi}{2}), \label{eq:g2}
\end{equation}
\labelz%
where ${\cal D}^{\frac{L}{2}}_{m^{\prime},m}(\frac{\pi}{2})$ is
determined by (\ref{eq:wg}).

Equations(\ref{eq:g1}--\ref{eq:g2}) are confined to the case of the
excitons initially  excited in one microcrystallite only.
Experimentally, we can also initially excite excitons in both
microcrystallites. If $P$ and $Q$ excitons are initially excited in
microcrystallites $A$ and $B$ then we have $j=(P+Q)/2$ and
$m=(P-Q)/2$, and so the initial state can be expressed as
$|\Phi(t=0)\rangle=
|P\rangle_{A}|Q\rangle_{B}=|\frac{P+Q}{2},\frac{P-Q}{2}\rangle_{s}$.
At time $t$, the wave function of the whole system can be written
as
\begin{eqnarray}
|\Phi(t)\rangle&=&\sum_{l^{\prime}=-(P+Q)/2}^{(P+Q)/2}
\exp\Big(\frac{t}{i\hbar}E_{\frac{P+Q}{2},l^{\prime}}\Big)\nonumber\\
&&\times
\langle\Psi_{\frac{P+Q}{2},l^{\prime}}|\frac{P+Q}{2},\frac{P-Q}{2}\rangle_s
|\Psi_{\frac{P+Q}{2},l^{\prime}}\rangle_{s}. \label{eq:pq0}
\end{eqnarray}
We can also obtain the entropy of entanglement for the above
initial condition as
\labela%
\begin{equation}
E^{(P,Q)}(t)=-\sum_{n^{\prime}={-(P+Q)}/2}^{(P+Q)/2}
|\beta_{n^{\prime}}^{\prime}|^{2}{\rm
ln}|\beta_{n^{\prime}}^{\prime}|^{2} \label{eq:pq1}
\end{equation}
with
\labelb%
\begin{equation}
\beta^{\prime}_{n^{\prime}}=\sum_{n=-(P+Q)/2}^{(P+Q)/2}{\cal
D}^{\frac{P+Q}{2}}_{\frac{P-Q}{2},n}(\frac{\pi}{2})e^{-\frac{t}{i\hbar}E_{\frac{P+Q}{2},n}}{\cal
D}^{\frac{P+Q}{2}}_{n^{\prime},n}(\frac{\pi}{2}). \label{eq:pq2}
\end{equation}
\labelz%
\vspace*{0mm}
Equations (\ref{eq:pq1}--\ref{eq:pq2}) for the entanglement were
obtained from the wave function (\ref{eq:pq0}) being a complete
solution of the model. Thus, the contributions of all elements
(also those off-diagonal) of the density matrix are included. In \
figure~\ref{fig2},  the entropies of the entanglement $E^{(P,Q)}$
are depicted as a function of $gt$ for different exciton numbers
$P$ and $Q$ in the first and second microcrystallite,
respectively, but for the same total number of five excitons in
both microcrystallites.  Numerical results show that the evolution
behavior of the entropy of the entanglement for $L=5$ is similar
to that for $L=2$. However, when we compare the maximum value of
the entropy of entanglement with the entropy of the maximally
entangled state, we see that a time $t$ at which the generated
entangled state is very close to a maximally entangled state
cannot be found for any $\chi/g$. The survival time of the maximum
entanglement is different for different initial conditions and
$\chi/g$. The smaller difference of initial populations
corresponds to longer time in a larger entanglement area for fixed
$\chi/g$ when the total number of excitons in two
microcrystallites is five.  We also find that the maximum
entanglement for fixed $\chi/g$ is the same for different initial
conditions for the total number $L$ of excitons in two
microcrystallites $L=2,3$ only, i.e.
$E_{\max}^{(2,0)}=E_{\max}^{(1,1)} =E_{\max}^{(0,2)}=\ln3$ and
$E_{\max}^{(3,0)} =E_{\max}^{(2,1)} =E_{\max}^{(1,2)}
=E_{\max}^{(0,3)} = 2 \ln2$. However, if $L>3$ then
$E_{\max}^{(L-n,n)}\neq E_{\max}^{(L-m,m)}$ for $n\neq m \leq
L/2$. Nevertheless, the differences are relatively small -  only
at the second digit after the decimal point, as we have checked
numerically up to ten excitons. Concluding, we find that the
maximum values of the entropy of entanglement depends both on the
initial conditions and on $\chi/g$ when $L>3$.

\begin{figure}
 \hspace*{2cm} \epsfxsize=60mm\epsfbox{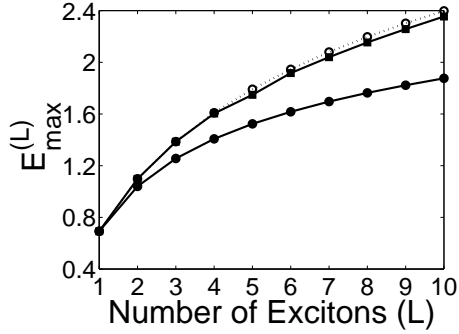}
\caption[]{The maximal values $E^{(L)}_{max}$ of the
time-dependent von Neumann entropy $E^{(L)}(t)$ are plotted as a
function of exciton number $L\leq10$ for $\chi/g=0$ (dots in solid
curve), $\chi/g=0.01$, $0.34$, and $0.8$ (squares in solid curve).
The values of the von Neumann entropy of the maximally entangled
states are marked by circles in the dotted curve for different
exciton numbers $L$.}\label{fig3}
\end{figure}

We can use equations (\ref{eq:g1}--\ref{eq:g2}) as an example to
further discuss the entanglement of the excitons between the two
subsystems $A$ and $B$ of the system for any initially given
exciton number. Here the entropy of the entanglement of two
subsystems is discussed up to ten excitons. An exact time $t$,
which corresponds to a maximally entangled state, can be obtained
when the system has one exciton. Numerically up to a evolution time
of $gt\leq 600$, when the total number of excitons in our system is
two or three, we find that the difference between the maximal
entropy of entanglement and the entropy of the maximally entangled
state is of the order $10^{-5}$ for $\chi/g\neq 0$. So, the
maximally entangled states can be approximately obtained. The
maximum values $E^{(L)}_{\rm max}$ of the entropy of the
entanglement for the two subsystems are also numerically calculated
by using (\ref{eq:g1})--(\ref{eq:g2}) up to the time $gt\leq 600$.
Figure \ref{fig3} shows a comparison of the maximal values
$E^{(L)}_{\rm max}$ of the entropy of the entanglement with the
values of the entropy of the entanglement for maximally entangled
states up to ten excitons for $\chi/g=0, \ 0.01, \ 0.34, \ 0.8$.
Figure \ref{fig3} shows that the entropies of the entanglement for
$\chi/g=0$  are smaller than those for $\chi/g=0.01, \ 0.34, \ 0.8$
for all exciton numbers, which means our system can achieve a
larger entanglement than that of the linear coupling model. At this
point, this coupled microcrystallites system can be considered as a
better source of entanglement with a fixed exciton number than that
of the linear coupling model with the same particle number. For a
fixed exciton number, we find that the difference in the maximum
values of the entropies of the entanglement between any two
different parameter ratios among $\chi/g=0.01, \ 0.34, \ 0.8$ is
very small. Every square in the solid curve of figure \ref{fig3}
actually denotes the maximum values of the entropies of the
entanglement for three different values of $\chi/g$.  By comparing
plots for a small ($\chi/g=0.01$) and a higher ($0.34$) values of
the nonlinearity in figures \ref{fig1}, \ref{fig2},  and
\ref{fig3}, it is clearly seen that the global maxima of
entanglement are practically independent of the ratio of $\chi/g$
(except $\chi/g$), but dependent of the initial conditions.
However, higher nonlinearity corresponding to a larger $\chi/g$
enables generation of globally higher entanglement for shorter
evolution times.

\section{Oscillations of exciton imbalance}

In the previous section, we show that the excitonic state of two
microcrystallites system has different degree of entanglement with
the  evolution. The population of the excitons in  the
microcrystallites  $A$ and $B$ is not at equilibrium. The
nano-technology opens up a new direction of research for direct
measurements of exciton dynamics.  Probing one exciton at a time
has been demonstrated by using the nonlinear
nano-optics~\cite{NH}.  It is possible for experimentalists to
observe the population of the excitons in the microcrystallites.
Considering that  experimentally, it is  easier to initially
excite excitons in one microcrystallite,  in this section our
discussions are the case when there are $L$ excitons excited
initially in one of the microcrystallites. We will show that the
excitonic population imbalance between two coupled
microcrystallites exhibits a collapse and revival phenomenon. The
time-dependent population difference of excitons between two
microcrystallites can be written as
\begin{equation}
\Delta N^{(L)}(t)=N^{(L)}_{1}(t)-N^{(L)}_{2}(t),
\end{equation}
where the average exciton number of each microcrystallite
$N^{(L)}_{l}(t)=\langle a^{\dagger}_{l}a_{l}\rangle$. The
evolution of the population difference of the excitons is similar
 for different initial conditions. We can discuss the
population imbalance of the two microcrystallites  using equation
(\ref{eq:T}), but here  we only consider the initial condition at
which there are $L$ excitons in microcrystallite $A$ and no exciton
in microcrystallite $B$. From (\ref{eq:13}), we can calculate the
number of  excitons in each microcrystallite as
\begin{eqnarray}
&&N^{(L)}_{l}(t)=\frac{L}{2}-(-1)^{l}\sum_{m^{\prime}=
-\frac{L}{2}}^{\frac{L}{2}} m^{\prime} \left |\beta_{m^{\prime}}
\right |^{2},
\end{eqnarray}
where $l=1$ represents the microcrystallite $A$ and $l=2$ denotes
microcrystallite $B$; time-dependent functions $\beta_{m^{\prime}}$
are given by equation (\ref{eq:g2}). Then we can obtain the
population difference of the excitons between the two
microcrystallites as
\begin{equation}
\Delta N^{(L)}(t)=2\sum_{m^{\prime}=-\frac{L}{2}}^{\frac{L}{2}}
m^{\prime} \left |\beta_{m^{\prime}} \right |^{2}.
\end{equation}
When only one exciton is initially excited in the microcrystallite
$A$,  the exciton population difference is
\begin{equation}
\Delta N^{(1)}(t)=\cos (2gt),
\label{eq:n1}
\end{equation}
which is a simple sinusoidal oscillation. The population difference
of the exciton is zero at times $gt=(2k+1)\frac{\pi}{4}$,  and
$\Delta N^{(1)}$ reaches a maximum at times $gt=k \frac{\pi}{2}$,
where $k$ is an integer. Comparing this result to equation
(\ref{eq:nnn}), we know that the two subsystems reach the maximal
entanglement when the exciton population difference of the two
microcrystallites is zero, but when the entropy of the entanglement
of the system is minimum, the population imbalance is maximum.

\begin{figure}[ht]
\hspace{15mm} \epsfxsize=12cm\epsfbox{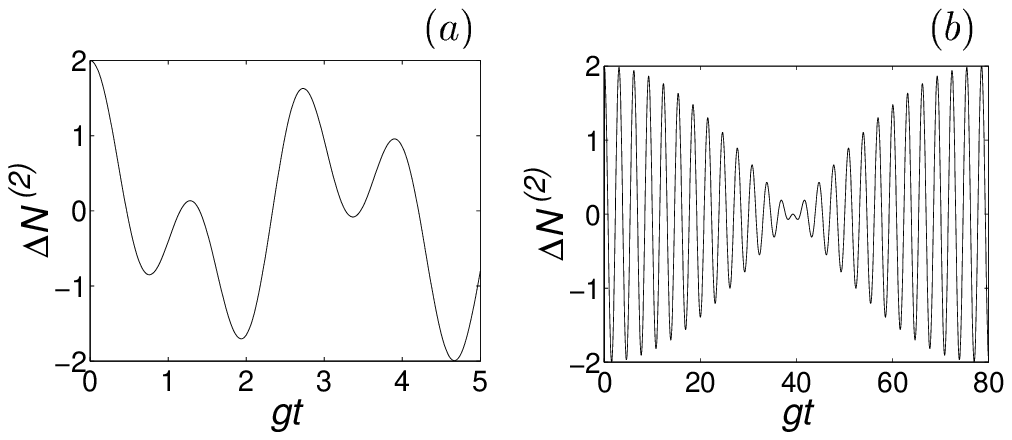} \caption[]{The
evolution of the population difference $\Delta N^{(2)}$ between two
microcrystallites is shown as a function of $gt$ for (a)
$\chi/g=0.34$ and (b) $\chi/g=0.01$.}\label{fig4}
\end{figure}

When there are two excitons initially excited in
microcrystallite $A$, we obtain
\begin{equation}
\Delta N^{(2)}(t)=2\cos (4\chi t)\cos[2gt +4\chi t]. \label{eq:N}
\end{equation}
Equation (\ref{eq:N}) indicates that the population difference
periodically oscillates with time. We numerically give the
evolution of the population difference as a function  of $gt$ in
figure \ref{fig4}.
Figure \ref{fig4} shows that the exciton population difference of
the two microcrystallites exhibits the beat effect. If the second
interaction between the excitons becomes weak,  the period of the
envelope becomes longer than that of the stronger  second
interaction between excitons.

\begin{figure}
\hspace{15mm} \epsfxsize=12cm\epsfbox{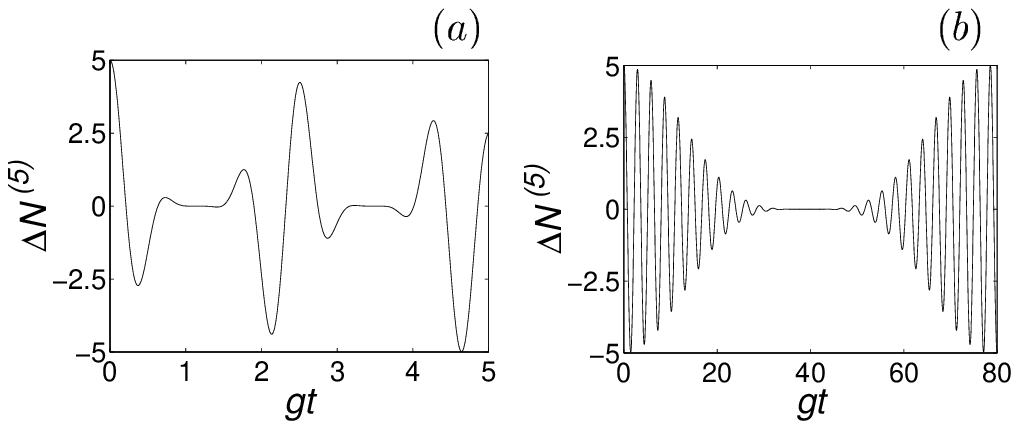} \caption[]{The
evolution of the population difference $\Delta N^{(5)}$ between two
microcrystallites is shown as a function of $gt$ for (a)
$\chi/g=0.34$, and (b) $\chi/g=0.01$.}\label{fig5}
\end{figure}

The general formula for the exciton population difference of two
microcrystallites can be given as
\begin{equation}
\Delta N^{(L)}(t)= L \cos^{L-1}(4\chi t)\cos [2gt +4(L-1)\chi t].
\end{equation}
 With increasing exciton number, the numerical results
show that exciton population difference evolution in time
resembles a collapse and revival phenomenon. As an example, the
 evolution of the exciton population difference in a system of
the coupled microcrystallites with a total number of $L=5$ excitons
is plotted in figure \ref{fig5} as a function of $gt$.  The smaller
is the ratio between the second coupling constant and the first
coupling constant, the clearer is this phenomenon.

\begin{figure}
 \hspace*{2cm} \epsfxsize=80mm\epsfbox{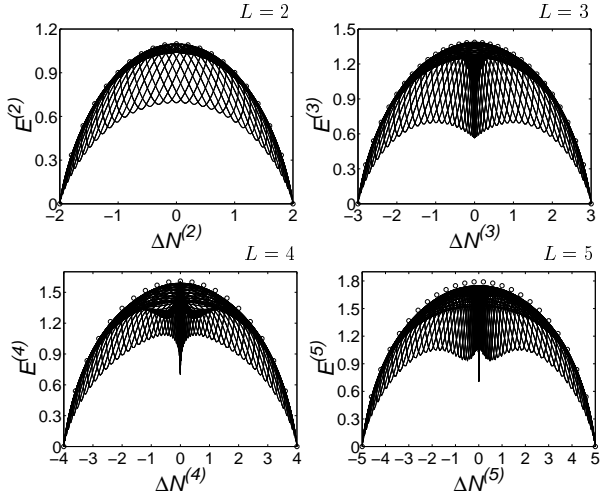}
\caption[]{The entropy of entanglement is plotted as a function of
the population difference of microcrystallites for $\chi/g=0.34$
when  the total number $L$ of excitons is 2, 3, 4, or 5. The
uppermost circles represent values estimated by the Jaynes
principle.}\label{fig6}
\end{figure}

Figure \ref{fig6} shows a relationship between the entropy of the
entanglement in our system and the exciton-population difference.
To estimate  the uppermost envelope functions of the curves in
figure \ref{fig6}, we apply the Jaynes principle of maximum
entropy~\cite{Jaynes}. In the Jaynes formalism, the entropy can
exclusively be expressed in terms of the mean values. The method
corresponds to averaging over the generalized grand canonical
ensemble of states satisfying  given constraints. Thus, the Jaynes
formalism of calculating the maximum entropy strongly resembles the
standard problems of entropy maximization in statistical mechanics
and thermodynamics. It is worth noting that, although the maximum
entropy principle ``has nothing to do with the laws of physics''
\cite{Jaynes}, it is a useful and widely applied  tool in physics,
e.g., in the quantum state reconstruction from data obtained in a
measurement of a physical process \cite{Buzek}. By applying the
Jaynes principle we find the maximum entropy to be expressed as a
function of the population difference as

\begin{eqnarray}\label{eq:Jay1}
&&E_{J}^{(L)}(\Delta N^{(L)})=-\sum_{n=0}^{L}p_{n}\ln
p_{n}=\frac{1}{2}(L+\Delta N^{(L)})\lambda ^{(L)}+\ln Z^{(L)},
\end{eqnarray}
where $p_{n}=\exp (-n\lambda ^{(L)})/Z^{(L)}\equiv x^{n}/Z^{(L)}$
are given in terms of Lagrange multiplier $\lambda ^{(L)}$ or
$x=\exp (-\lambda ^{(L)}) $ and the generalized partition function

\begin{eqnarray}\label{eq:Jay2}
Z^{(L)}=\frac{x^{L+1}-1}{x-1}.
\end{eqnarray}
The Lagrange multiplier $\lambda ^{(L)}$ can be calculated from

\begin{eqnarray}\label{eq:Jay3}
\Delta N^{(L)}=f(x)=L+\frac{2}{1-x}-2\frac{1+L}{1-x^{L+1}}.
\end{eqnarray}
To obtain an explicit formula for $E_{J}^{(L)}$ one has to find the
relation inverse to $\Delta N^{(L)}=f(x)$. By closer inspection of
(\ref{eq:Jay3}), we conclude that the analytical solutions of
$x=f^{-1}(\Delta N^{(L)})$ exist only for $L$=2, 3, and 4. For
example,  the desired solution of $x=f^{-1}(\Delta N^{(2)})$ for
$N=2$ is given by

\begin{equation}\label{eq:Jay4}
x\equiv \exp (-\lambda ^{(2)})=\frac{\Delta
N^{(2)}+\sqrt{16-3(\Delta N^{(2)})^{2}}}{2(2-\Delta N^{(2)})}.
\end{equation}
For brevity, we do not present our  analytical, but rather
complicated, solutions  for $L=3$ and $4$. The solutions for $L\geq
5$ are to be found numerically. The curves marked by circles in
figure 6 correspond to our analytical ($L=2, 3, 4$) and numerical
($L=5$) estimations of the maximum entropy of entanglement as a
function of $\Delta N^{(L)}$. It is seen, in agreement with our
analysis of figure 3, that with the increasing number $L$ of
excitons, the time-maximized  entanglement $E^{(L)}_{\max}(t)$,
which can be generated in our system, increases in comparison to
the lower-exciton systems $E^{(L)}_{\max}(t)> E^{(L-1)}_{\max}(t)$
for the same population difference $\Delta N\equiv \Delta
N^{(L)}=\Delta N^{(L-1)}\neq \pm $ $L$ (with any $L$) for a given
$\kappa/g$. The states described by $E_{J}^{(L)}(\Delta N^{(L)})$
for $\Delta N=0$ correspond to the maximally entangled states. The
numerical results depicted in figure \ref{fig6} show that the
maximum entropy of the entanglement always corresponds to the
points of the exciton-population balance. However, the condition of
the exciton-population balance does not always imply the maximum
entropy of the entanglement. For any population-imbalance points,
the entropy of the entanglement is smaller, and when the population
difference reaches its maximum, the entropy of the entanglement
vanishes. \vspace{0.5cm}

\section{Conclusions}

We have discussed the entanglement of the excitonic states in the
system of the coupled microcrystallites with a fixed total exciton
number by the entropy of the entanglement. When the total number
of excitons is one, the maximally entangled state can exactly be
obtained. The maximally entangled states can approximately  be
generated for two or three excitons. However, when the exciton
number is more than three, we cannot obtain the maximally
entangled state, but it is observed that the nonlinear interaction
between excitons increases the maximum values of the entropy of
the entanglement to higher values  than those of the other models
based on the linear coupling. So this exciton system can be used
as a good source of entanglement. We also find that the
entanglement between the two microcrystallites depends on the
initial conditions, the different initial conditions will result
in different entanglements.

The oscillations of exciton population between two
microcrystallites have also been investigated analytically and
numerically. If the system is prepared with one exciton, the
population imbalance between two microcrystallites reaches the
maximum when the entanglement of the two microcrystallites is the
minimum; and the population reaches the balance when the entropy of
the entanglement of the system is maximum. But when the exciton
number is more than two, the numerical results showed that we could
find the maximum entropy of the entanglement at the points of the
exciton population balance. At some larger population imbalance
points, the entropy of the entanglement was found to be smaller.
The reason is that when the population imbalance becomes the
largest, all excitons are concentrated on one  microcrystallite,
then the entanglement between two microcrystallites decreases. But
in the collapse area, the average population of each
microcrystallite is equal, so it is possible to find the maximal
entropy of the entanglement. We hope these phenomena  can be
observed by experimentalists with the development of the
nano-technology.

The main conclusion of our paper is that nonlinear interactions,
resulting from high exciton density between microcrystallites, can
increase exciton entanglement in comparison to a model of
linearly-coupled microcrystallites with lower exciton density.  We
believe that this result could be important for
semiconductor-based quantum information processing. Our conclusion
is not only valid for the studied system, but can also be
generalized to other Kerr nonlinear vs linear interaction models.

\section*{Acknowledgments}
Authors are grateful to Makoto Kuwata-Gonokami and Yuri P. Svirko
for  helpful discussions. Yu-xi Liu acknowledges the support of
Japan Society for the Promotion of Science (JSPS).



\section*{References}

\begin{thebibliography}{99}

\bibitem{Zeilinger} Bouwmeester D, Pan J-W, Daniel M,
Weinfurter H and Zeilinger A 1997 {\em Nature (London)} {\bf 390}
575

\bibitem{B} Furusawa A, S\o rensen J I, Braunstein S L, Fuchs C A,
Kimble H J and Polzik E S 1998 {\em Science} {\bf 282} 706

\bibitem{mmm} Duan L M, Giedke G, Cirac J I and Zoller P
2000 {\em Phys. Rev. Lett.} {\bf 84} 4002

\par\item[] Duan L M, Giedke G, Cirac J I and Zoller P
2000 {\em Phys. Rev. A} {\bf 62} 032304

\bibitem{c} Cochrane P T, Milburn G J and Munro W J
2000 {\em  Phys. Rev. A} {\bf 62} 062307

\bibitem{4} Barenco A, Deutsh D, Ekert A and Jozsa R 1995
{\em Phys. Rev. Lett.}  {\bf 74} 4083

\par\item[] Loss D and DiVincenzo P 1998
{\em Phys. Rev. A }{\bf 57} 120

\par\item[] Imamo\v{g}lu A, Awschalom D D, Burkard G, DiVincenzo D P,
Loss D, Sherwin M and Small A 1999 {\em Phys. Rev. Lett.}  {\bf
83} 4204

\bibitem{44} Oh J H, Ahn D and Hwang S W 2000 {\em Phys. Rev. A} {\bf 62}
052306

\bibitem{5} Chen G, Bonadeo N H, Steel D G, Gammon D,
Katzer D S, Park D and Sham L J 2000 Science {\bf 289} 1906

\bibitem{6} Bayer M, Hawrylak P, Hinzer K, Fafard S,
Korkusinski M, Wasilewski Z R, Stern O and Forchel A 2001 Science
{\bf 291} 451

\bibitem{7} Quiroga L and Johnson N F 1999
{\em Phys. Rev. Lett.} {\bf 83} 2270

\par\item[] Reina J H, Quiroga L and Johnson N F 2000 {\em Phys. Rev. A}
{\bf 62} 012305

\bibitem{99} Yi X X, Jin G R, Zhou D L 2001 {\em Phys. Rev. A} {\bf 63}
062307

\bibitem{yu-xi2}
Miranowicz A, \"Ozdemir \c{S} K, Liu Yu-xi, Koashi M, Imoto N and
Hirayama Y 2002 {\em Phys. Rev. A}  {\bf 65} 062321

\bibitem{wang}Wang X, Feng M and Sanders B C 2003 {\em Phys.
Rev. A} {\bf 67} 022302

\bibitem{3} Livermore C, Crouch C H, Westervelt R M,
Campman K L and Gossard A C 1996 Science {\bf 274} 1332

\bibitem{hirayama} Fujisawa T, Oosterkamp T H, van der Wiel W G,
Broer B W, Aguado R, Tarucha S, Kouwenhoven L P 1998 Science {\bf
282} 932

\par\item[] Oosterkamp T H, Fujisawa T, van der Wiel W G, Ishibashi K,
Hijman R V, Tarucha S and Kouwenhoven L P 1998 Nature {\bf 395}
873

\bibitem{Han} Hanamura E 1988 Phys Rev B {\bf 38} 1228

\bibitem{Ban} Belleguie L and B\'anyai L 1991 {\em Phys. Rev. B}
{\bf 44} 8785

\par\item[] B\'anyai L and Koch S W 1993 {\em Semiconductor Quantum Dots}
(Singapore: World Scientific)

\bibitem{Eng} Engelmann A, Yudson V I and Reineker P 1998
{\em Phys. Rev. B} {\bf 57} 1784

\bibitem{yu-xi}
Liu Yu-xi,  \"Ozdemir S K, Koashi M and Imoto N 2002 {\em Phys.
Rev. A} {\bf65} 042326

\par\item[] Liu Yu-xi, Miranowicz A, Koashi M and Imoto N 2002 {\em Phys.
Rev. A} {\bf 66} 062309

\par\item[] Liu Yu-xi, Miranowicz A, \"Ozdemir S K, Koashi M and Imoto N
2003 {\em Phys. Rev. A} {\bf 67} 034303

\bibitem{d} Chernyak V and Mukamel S 1996 {\em J. Opt. Soc. Am. B}
{\bf 13} 1302

\bibitem{Sch65} Schwinger J 1965, US Atomic Energy Commission Report
No. NYO-3071 (US GPO, Washington, D.C.); reprinted in Biedenharn L
C and van Dam H (eds) 1965 {\em Quantum Theory of Angular Momentum}
(New York: Academic) p 229

\bibitem{Sak94}Sakurai J J 1994 {\em Modern Quantum Mechanics}
(New York: Addison-Wesley)

\bibitem{Ben96a}Bennett C H, Bernstein H J, Popescu S, and Schumacher B 1996 {\em
Phys. Rev. A} {\bf 53} 2046

\bibitem{Ben96b}Bennett C H, DiVincenzo D P, Smolin J A, and Wootters W K 1996
{\em Phys. Rev. A} {\bf 54} 3824

\bibitem{Ved98}Vedral V and Plenio M B 1998 {\em Phys. Rev. A} {\bf 57} 1619

\bibitem{Pho91}Phoenix S J D and Knight P L 1991 {\em Phys. Rev. A} {\bf 44}
6023; 1991 {\em Phys. Rev. Lett.} {\bf 66} 2833; 1988 {\em Ann.
Phys.} {\bf 186} 381

\bibitem{NH} Bonadeo N H, Chen Gang, Gammon D, Katzer D S,
Park D and Steel D G 1998 {\em Phys. Rev. Lett.}  {\bf 81} 2759

\bibitem{Jaynes}
Jaynes E T 1957 {\em Phys. Rev.} {\bf 106} 620; {\bf 108} 171

\bibitem{Buzek}  Bu\v{z}ek V, Adam G and Drobn\'{y} G 1996
{\em Phys. Rev. A} {\bf 54} 804

\end{thebibliography}
\end{document}